\documentclass[12pt]{article}
\usepackage{graphicx}
\usepackage{amsmath}
\usepackage{tabularx}
\usepackage{float}
\usepackage{geometry}
\usepackage{setspace}
\usepackage[utf8]{inputenc}
\usepackage{cite}  
\usepackage{xcolor}
\usepackage{url} 
\usepackage{siunitx}
\usepackage{orcidlink}
\DeclareSIUnit{\au}{AU}
\DeclareSIUnit{\mas}{mas}
\DeclareUnicodeCharacter{03B1}{$\alpha$}

\title{Stellar Parameters and Evolutionary Pathways of the Subgiant system HIP 72217}
\author{Naufa Nazar $^{1,\ast}$ \orcidlink{0009-0006-2590-0752}, Mashhoor A. Al-Wardat$^{1,2,\ast}$ \orcidlink{0000-0002-1422-211X},Ahmad Abushattal$^{3}$ \orcidlink{0000-0002-7796-6562}, \\ Hassan B. Haboubi$^{1,\ast}$ \orcidlink{0009-0004-7254-0724}}

\date{}

\begin{document}

\maketitle

\noindent
$^{1}$ \textit{Department of Applied Physics and Astronomy, College of Sciences, University of Sharjah, P.O.Box 27272 Sharjah, United Arab Emirates} \\
$^{2}$ \textit{Sharjah Academy for Astronomy, Space Sciences and Technology, University of Sharjah, P.O.Box 27272 Sharjah, United Arab Emirates} \\
$^{3}$ \textit{Department of Physics, Al-Hussein Bin Talal University, P. O. Box 20, Ma’an 71111, Jordan} \\
$^{\ast}$ \textit{Corresponding Authors:} 
\texttt{<u22200624@sharjah.ac.ae>},
\texttt{<u23103604@sharjah.ac.ae>}, \texttt{<malwardat@sharjah.ac.ae>}\\

\texttt{Jordan Journal of Physics https:\url{//jjp.yu.edu.jo/index.php/jjp/index}: Received on 15 April 2025, Accepted on 18 May 2025}

\section*{Abstract}
In this study, we applied Al-Wardat's method to analyze the subgiant system HIP72217 for which we obtained accurate parameters including stellar masses, effective temperatures ($T_{\text{eff}}$) and system age.For the primary component we determined a stellar mass of  $M_A = 1.14 \pm 0.15\,M_{\odot}$ and effective temperature $T_{\text{eff,A}} = 6125 \pm 50$\,K while for the secondary component we obtained the values of  $M_B = 1.12 \pm 0.14\,M_{\odot}$ and $T_{\text{eff,2}} = 5950 \pm 50$\,K. The system's age was estimated to be  $3.548 Gyr$, which is consistent with the predicted evolutionary period of a subgiant binary. The evolutionary timeline of HIP\,72217 becomes clearer through our study, which also demonstrates Al-Wardat's approach as an effective approach for binary star system characterization. These findings contribute to a better understanding of the physical mechanisms that control subgiant binary evolution and their broader role in stellar evolutionary processes.

\textbf{Keywords---} Binary Stars, Al-Wardat's Method, Subgiant, Evolutionary Tracks, Isochrones. 

\section{Introduction}
The study of stellar star systems leads to enhanced understanding of stellar evolution because researchers can directly measure essential stellar properties including mass, radius, and luminosity \cite{Torres_2009, docobo2017precise, abushattal2019most, alnaimat2023jewel, abushattal2022astroinformatics}. Subgiant stars serve as essential research objects to study the main-sequence to giant-phase transition because they reveal information about core hydrogen exhaustion and shell hydrogen-burning processes \cite{2006epbm.book.....E}. The study of subgiants in binary or triple systems delivers exceptional value since their evolutionary pathways become directly impacted by the dynamical interactions and mass transfer between components. The subgiant system HIP~72217 provides researchers with an exceptional chance to study these phenomena extensively.\\

Binary systems fall into three main types: astrometric binaries, spectroscopic binaries, and eclipse binaries. One technique alone cannot determine an individual system mass. Scientists worldwide have used speckle interferometry for fifty years to observe binary stars effectively and accurately \cite{abushattal202424,docobo2018iau,abushattal2023advancements,hussein2022atmospheric}. Due to the stages of star formation, multiple stars usually contain the most massive exoplanets. There are many systems in which binary stars host giant exoplanets. A binary star's physical properties determine the habitability and stability of an exoplanet, both of which are essential parameters in exoplanet studies. This study mainly focuses on parameters such as individual masses, semi-major axes, luminosities, and temperatures\cite{abushattal2017modeling,alnaimat12025jewel,abushattal2019extrasolar,algnamat2022precise,alameryeen2022physical,abushattal2022exoplanets}. Al-Wardat’s method for analyzing stellar systems is a useful tool for determining accurate stellar parameters. This method combined photometry and model atmospheres to create synthetic spectral energy distribution (SED) and to identify sub-components or sub-systems of the components of the stellar system. As it has been mentioned earlier, Al-Wardat’s method enables the determination of the masses, radii, effective temperatures, and luminosities with high accuracy when the observed data is compared to the theoretical models (see: \cite{Yousef2021, Al-Wardat2003,Al-Wardat2007, Masda2016,2023AJ....165..221A, Masda2018,al2017physical, al2021precise, Masda2021, Masda2024}). 
HIP 72217 is a triple star system located in the Libra constellation, It is composed of a close inner binary pair (HIP 72217 A and B) with a separation of 0.182 mas orbited by a more distant companion (HIP 72217 C) with a separation of 6.923 mas. Applying this to a system such as HIP72217 helps to gain insight into the physical characteristics and possible evolutionary stages of stellar systems, especially those in the subgiant category.\\

Although the method provides an effctive analysis of the stellar parameters, it also comes with certain limitations. The main drawback of this method stems from its requirement of theoretical stellar atmosphere models such as Kurucz or PHOENIX which depend on assumptions regarding static, plane-parallel and local thermodynamic equilibrium (LTE) conditions. Stellar atmosphere models used in this method do not provide comprehensive simulation of non-LTE phenomena along with stellar activity features or convective overshooting effects that are significant in evolved or active stars. The method requires complete knowledge of stellar components because any unresolved objects or interstellar effects not properly handled introduce systematic measurement errors. The precision of synthetic spectral energy distribution (SED) fitting depends largely on the initial values of metallicity and interstellar reddening because incorrect estimates from these parameters affect the determination of temperature and luminosity and radius. Al-Wardat’s method needs external verification through independent methods such as interferometry and spectroscopy and asteroseismology to eliminate parameter uncertainty and validate results.

\begin{figure}[H]
    \centering
    \includegraphics[width=0.9\textwidth]{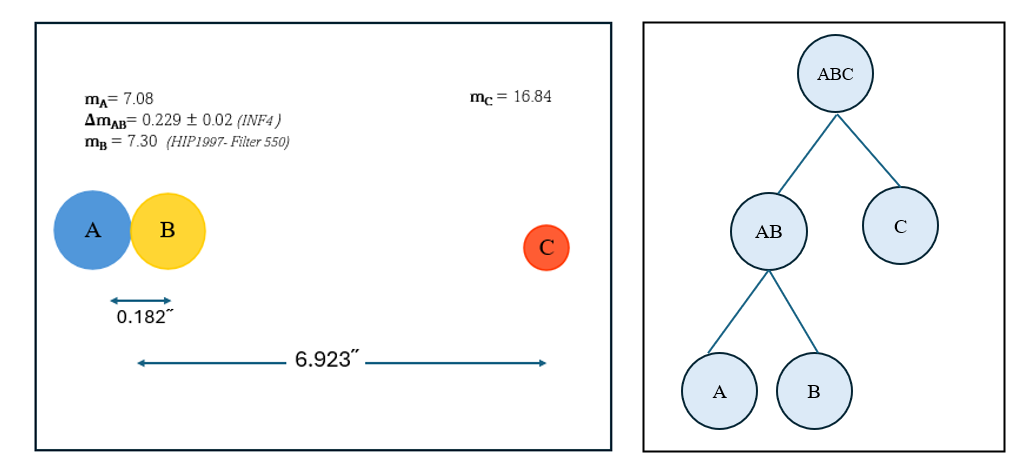}
    \caption{Left: Pictorial representation of the three components A, B and C of the system HIP 72217 with their magnitudes( $m_{vA}$=7.08,  $m_{vB}$ = 7.30,  $m_{vC}$=16.84) and magnitude differences\cite{ESA1997} Their separation in mas is also denoted in the picture.
    Right: Hierarchical chart showing the combinations of the three components.}
    \label{fig:hierarchy}
\end{figure}

\section{Observational Data}
The study of the HIP72217 system involved gathering observational data from multiple sources to analyze its stellar components. Table~\ref{tab:hip72217_Obs} presents the observational data, which includes apparent magnitudes in the B and V filters, spectral classifications, the $m_{\text{V}}$ magnitude difference, and the parallax measurements derived from Hipparcos.

\begin{table}[H]
    \centering
    \caption{Observational Data for HIP72217}
    \begin{tabular}{lcll}
    \hline
    \multicolumn{4}{c}{}\\[-1em]
    & \textbf{HIP 72217} & & \textbf{Source} \\
    & \textbf{HD 129980} & & \\
    \hline
    $\alpha_{2000}$ & 14h 46m 10.89 & & Simbad\cite{SIMBAD} \\
    $\delta_{2000}$ & $-21^\circ\,10'\,33.46$ & &Simbad\cite{SIMBAD}\\
    Sp.\ Type & G1V & & Simbad \cite{SIMBAD}\\
    $(\Delta m_v)A,B$ & $0.229 \pm 0.02$ & & INF4 \cite{Fourth}\\
    $(\Delta m_v)AB,C$ & $10.41 \pm 0.10$ & & UMSC \cite{UMSC}\\
    $A_v$ & 0.005 & & G-Tomo \cite{EXPLORE} \\
    $V_J$ & $6.43 \pm 0.01$ & & Tycho\cite{2003yCat.2246....0C} \\
    $B_J$ & $7.04 \pm 0.02$ & & Tycho\cite{2003yCat.2246....0C} \\
    $(B - V)_J$ & $0.603 \pm 0.004$ & &Tycho\cite{2003yCat.2246....0C} \\
    $\pi_{\mathrm{Hip}}$ & $23.64 \pm 1.07$ & &ESA1997 \cite{1997ESASP1200.....E} \\
    $\pi_{\mathrm{Hip}}$ & $24.24 \pm 0.63$ & &VNHR2007  \cite{2007ASPC..364.....S} \\
    \hline
    \end{tabular}
    \label{tab:hip72217_Obs}
\end{table}

\section{Spectrophotometric Analysis of the System}
We applied \textbf{Al-Wardat's method} for analyzing stellar systems to study the target system. This technique combines theoretical stellar atmospheres with photometric observations to derive precise physical parameters(see: \cite{ Al-Wardat2007, Al-Wardat2014c, Al-Wardat2016}). The analysis starts with the magnitude difference ($\Delta m$) between components and the system's total visual magnitude ($m_v$). The spectral energy distribution (SED) is then constructed using the effective temperature and surface gravity ($\log g$) (see: \cite{Al-Wardat2002,Al-Wardat2009,Al-Wardat2014a}), where the flux ratio of the components is given by:

\begin{equation}
\frac{f_{1}}{f_{2}} = 2.512^{-\Delta m}
\end{equation}

\begin{equation}
m_v = -2.5\log(f_{1} + f_{2})
\end{equation}

From these, we determine the apparent magnitudes of both stars:

\begin{equation}
m_{v}^{(A)} = m_{v} + 2.5 \log\bigl(1 + 10^{-0.4\,\Delta m}\bigr)
\end{equation}

\begin{equation}
m_{v}^{(B)} = m_{v}^{(A)} + \Delta m 
\end{equation}

The uncertainties in these magnitudes are calculated as:

\begin{equation}
\sigma^{2}_{m_{v}^{(A)}} = \sigma^{2}_{m_{v}} + \biggl(\frac{10^{-0.4\,\Delta m}}{1 + 10^{-0.4\,\Delta m}}\biggr)^{2} \sigma^{2}_{\Delta m}
\end{equation}

\begin{equation}
\sigma^{2}_{m_{v}^{(B)}} = \sigma^{2}_{m_{v}^{(A)}} + \sigma^{2}_{\Delta m}
\end{equation}

Next, the absolute magnitudes ($M_v$) are derived using the distance ($d$ in parsecs) and extinction ($A_v$):

\begin{equation}
M_{v} = m_{v} + 5 - 5\log(d) - A_{v}
\end{equation}

The errors in absolute magnitudes are:

\begin{equation}
\sigma^{2}_{M_{v}^{(A)}} = \sigma^{2}_{m_{v}^{(A)}} + \bigl(\ln(0.2\pi)\bigr)^{2}\sigma^{2}_{\pi}
\end{equation}

\begin{equation}
\sigma^{2}_{M_{v}^{(B)}} = \sigma^{2}_{m_{v}^{(B)}} + \bigl(\ln(0.2\pi)\bigr)^{2}\sigma^{2}_{\pi}
\end{equation}

Using these, we estimate the effective temperature and bolometric correction (BC) from \cite{1992adps.book.....L, Gray}, then compute the bolometric magnitude:

\begin{equation}
M_{\mathrm{bol}} = M_{v} + \text{BC}
\end{equation}

The stellar luminosity follows from:

\begin{equation}
M_{\mathrm{bol}} - M_{\odot,\mathrm{bol}} = -2.5 \log\biggl(\frac{L}{L_{\odot}}\biggr)
\end{equation}

The radii are then determined via:

\begin{equation}
\log\biggl(\frac{R}{R_{\odot}}\biggr) = 0.5 \log\biggl(\frac{L}{L_{\odot}}\biggr) - 2\log\biggl(\frac{T}{T_{\odot}}\biggr)
\end{equation}

Masses are estimated using empirical relations from \cite{1992adps.book.....L}, and the surface gravity is found using:

\begin{equation}
\log(g) = \log\biggl(\frac{M}{M_{\odot}}\biggr) - 2\log\biggl(\frac{R}{R_{\odot}}\biggr) + 4.43
\end{equation}

To verify the synthetic SEDs, \textbf{Al-Wardat's method} employs synthetic photometry, where magnitudes are calculated as:

\begin{equation}
    m_p = -2.5 \log \frac{\int P_p(\lambda) F_{\lambda, s}(\lambda) \, d\lambda}{\int P_p(\lambda) F_{\lambda, r}(\lambda) \, d\lambda} + ZP_p,
\end{equation}

where:
\begin{itemize}
    \item $m_p$: synthetic magnitude in passband $p$,
    \item $P_p(\lambda)$: normalized sensitivity function of passband $p$,
    \item $F_{\lambda, s}(\lambda)$: synthetic SED of the star,
    \item $F_{\lambda, r}(\lambda)$: Vega's reference SED,
    \item $ZP_p$: Vega-based zero-point calibration \cite{2007ASPC..364.....S}.
\end{itemize}

Color indices are computed, and SED magnitudes are estimated using Al-Wardat’s synthetic photometry method. To achieve the best fit between synthetic and observational visual magnitudes, color indices, and magnitude differences, an iterative method is employed (see: \cite{Yousef2021, Al-WardatBarstow2021, Al-Wardat2017,Al-Wardat2021} ).\\

\section{Orbital Analysis}
%\label{subsec:orbital_solution}

The orbital solution for the HIP 72217 system was derived using Tokovinin's ORBITX code \cite{2016AJ....151..153T}. This method requires the following input parameters:
\begin{itemize}
    \item Right Ascension (RA) in the format HH.MMSS.
    \item Declination (Dec) in the format DD.MMSS.
    \item Orbital period \( P \) (initial value in years).
    \item Epoch of periastron passage \( T \) (in years).
    \item Eccentricity \( e \).
    \item Semi-major axis \( a \) (in arcseconds).
    \item Position angle of the line of nodes \( \Omega \) (in degrees).
    \item Longitude of periastron \( \omega \) (in degrees).
    \item Inclination \( i \) (in degrees).
    \item Semi-amplitudes of radial velocities \( K_1 \) and \( K_2 \) (in km/s) for the primary and secondary components, respectively.
    \item Systemic velocity \( V_0 \) (in km/s).
\end{itemize}

These parameters were sourced from the Sixth Catalog of Orbits of Binary Stars \cite{wds_orb6}. Additionally, angular separations \( \rho \) (in arcseconds) and position angles \( \theta \) (in degrees) were obtained from the Fourth Catalog of Interferometric Measurements of Binary Stars \cite{Fourth}.

The dynamical mass of the system was calculated using Kepler's Third Law for binary stars:
\begin{equation}
    M_{\text{dyn}} = M_A + M_B = \left( \frac{a^3}{\pi^3 P^2} \right) M_{\odot},
    \label{eq:kepler_third_law}
\end{equation}
where \( \pi \) is the parallax, a is the semi-major axis of the relative orbit of the binary system in arcseconds, and P is the orbital period in years. 

The formal error in the mass was determined using:
\begin{equation}
    \frac{\sigma_M}{M} = \sqrt{9 \left( \frac{\sigma_\pi}{\pi} \right)^2 + 9 \left( \frac{\sigma_a}{a} \right)^2 + 4 \left( \frac{\sigma_P}{P} \right)^2}.
    \label{eq:mass_error}
\end{equation}

\section{Results}
\subsection{Spectrophotometric Analysis}
\begin{figure}[H]
    \centering
    \includegraphics[width=0.9\textwidth]{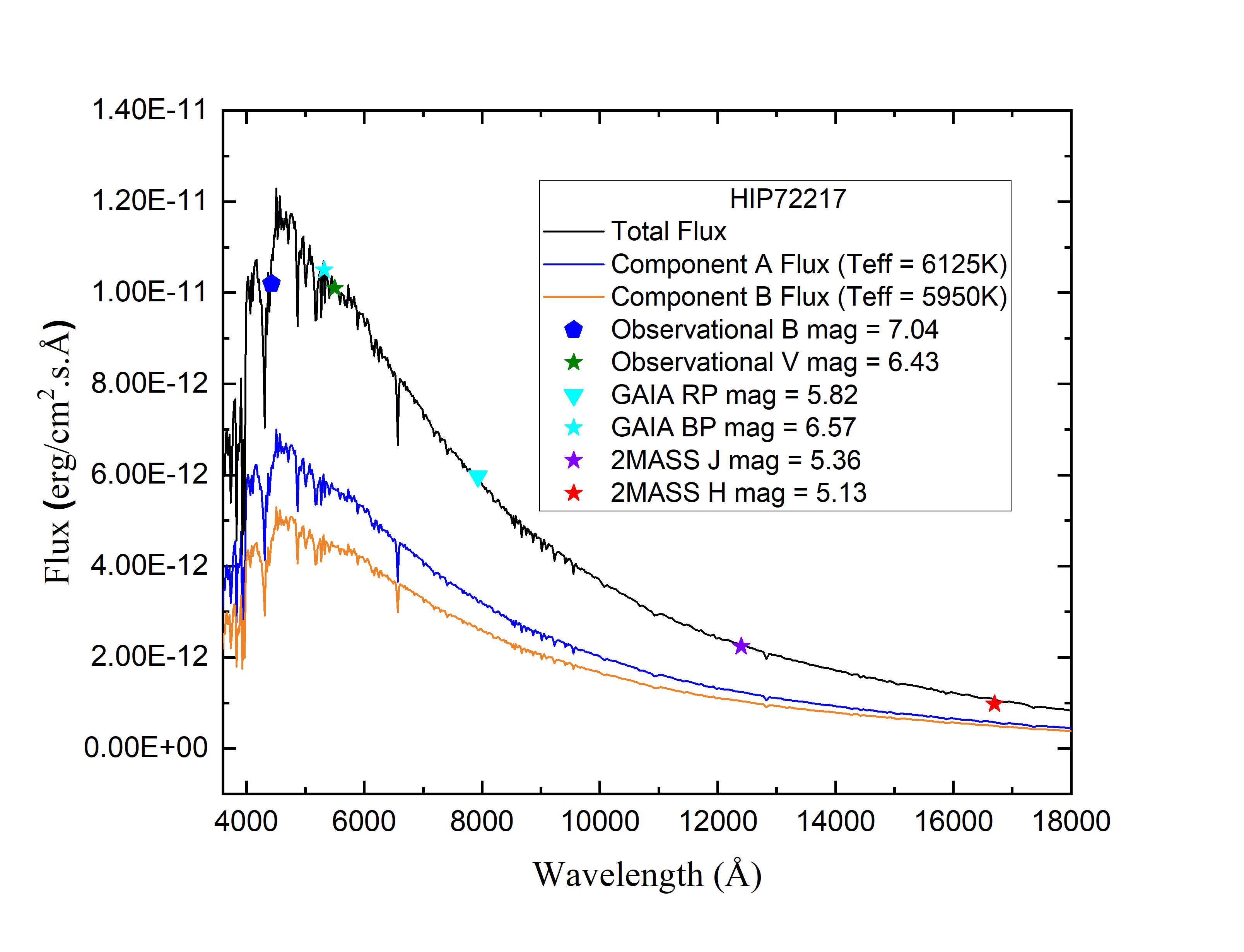}
    \caption{Spectral Energy Distribution (SED) of HIP 72217, showing the combined synthetic SED of the binary system along with individual flux contributions from each component. Observed magnitudes from Gaia\cite{2022yCat.1355....0G}, 2MASS\cite{2003yCat.2246....0C} and Tycho\cite{2002A&A...384..180F} are overlaid for comparison.}
    \label{fig1}
\end{figure}

%\subsection{Al-Wardat's Method}
%The Spectral Energy Distribution (SED) of HIP 72217, reconstructed using synthetic photometry, is shown in 
Figure \ref{fig1} shows the observed photometric data points from various bands plotted against the synthetic SED generated for the best-fitting stellar parameters of the two components. 

The individual contributions of Component A and Component B to the total flux are also displayed, highlighting their relative influence on the combined spectrum. The agreement between the observed photometric data points (colored markers) and the model SED (black curve) demonstrates the robustness of the derived stellar parameters, particularly the effective temperatures ($T_{\text{eff}, A} = 6125$ K, $T_{\text{eff}, B} = 5950$ K) and surface gravities ($\log g_A = 4.14$, $\log g_B = 4.20$).\\

The synthetic magnitudes derived from this SED fitting align well with the observed magnitudes, with minimal deviations in the Gaia and Johnson-Cousins photometric bands. This provides additional validation of the accuracy of the modeled stellar properties.\\

Using the SEDs and Equation 14, we calculated the synthetic magnitudes (This work) of the total system and the components, as listed in Table~\ref{tab:sed_magnitudes}. And the difference in magnitude $\Delta mv$ between both components is recorded in Table~\ref{tab:delta_mv_comparison} where the synthetic value is compared to the observed. The results we got match the observational values recorded in Table~\ref{tab:hip72217_Obs}. This strong correlation can be considered as a kind of confirmation of our approach because the parameters estimated in the analysis are consistent with theoretical expectations and with the observed values of the star. This correspondence further supports the validity of the method and confirms the credibility of the determined stellar parameters seen in Table~\ref{tab:parameters}. 
We included in our analysis of the binary system AB a third component C, which is known to have a mass of 0.17~M$_\odot$. Nevertheless, our study was mostly based on the AB components, since there were no observational parameters, such as the $B-V$ color index. We had taken the effect of component C into account in the overall system dynamics, but due to lack of data we were unable to do a detailed analysis of the third body. However, we were able to estimate its effective temperature ($T_{\text{eff}}$) of 3250~K and radius of 0.28~R$_\odot$, and a synthetic magnitude that agrees with the observed value of 16.84. Using this estimation and the observed $\Delta mv$(AB,C) we were able to conduct some analyses for the component.\\

\begin{table*}[h!]
    \centering
    \small % reduce font size slightly
    \caption{Synthetic (Syn) magnitudes of components A, B, and C, and the total flux compared with available observed (Obs) magnitudes.}
    \label{tab:sed_magnitudes}
    \begin{tabularx}{\textwidth}{l *{5}{>{\centering\arraybackslash}X}}
    \hline
    \textbf{Filter} & \textbf{Syn Total ($\sigma = 0.05$)} & \textbf{Obs Total} & \textbf{Syn Star A ($\sigma = 0.02$)} & \textbf{Syn Star B ($\sigma = 0.04$)} & \textbf{Syn Star C} \\
    \hline
    B Johnson & 7.04 & $7.04 \pm 0.02$ & 7.64 & 7.96 & 18.7 \\
    V Johnson & 6.42 & $6.43 \pm 0.01$ & 7.06 & 7.33 & 16.82 \\
    R Cousins & 5.81 & -- & 6.44 & 6.69 & 15.3 \\
    I Cousins & 5.43 & -- & 6.08 & 6.31 & 14.19 \\
    J 2MASS   & 5.37 & $5.36 \pm 0.027$ & 6.03 & 6.22 & 12.55 \\
    H 2MASS   & 5.04 & $5.10 \pm 0.04$ & 5.67 & 5.84 & 11.38 \\
    G Gaia    & 6.31 & $6.30 \pm 0.001$ & 6.95 & 7.20 & 15.44 \\
    $B_\mathrm{p}$ Gaia & 6.64 & $6.57 \pm 0.001$ & 7.26 & 7.55 & 17.31 \\
    $R_\mathrm{p}$ Gaia & 5.85 & $5.82 \pm 0.001$ & 6.49 & 6.72 & 14.29 \\
    $B - V$ & 0.602 & 0.603 & 0.583 & 0.627 & 1.877 \\
    $B_\mathrm{p} - R_\mathrm{p}$ & 0.797 & 0.754 & 0.769 & 0.832 & 3.06 \\
    \hline
    \end{tabularx}
\end{table*}

\begin{table}[h!]
\centering
\small
\caption{Comparison of Synthetic and Observed $\Delta m_V$(A,B) values}
\label{tab:delta_mv_comparison}
\begin{tabularx}{\linewidth}{l>{\centering\arraybackslash}X>{\centering\arraybackslash}X}
\hline
\textbf{Type} & $\boldsymbol{\Delta m_V}$ & \textbf{Source} \\
\hline
Synthetic   & $0.26 \pm 0.05$ & (This Work) \\
Observed 1  & $0.20 \pm 0.05$ & Tok2010 S~\cite{Tokovinin2010} \\
Observed 2  & $0.20 \pm 0.04$ & Tok2012d St~\cite{Tokovinin2014a} \\
Observed 3  & $0.30 \pm 0.06$ & Tok2015c St~\cite{Tokovinin2015c} \\
Observed 4  & $0.30 \pm 0.05$ & Tok2016a St~\cite{Tokovinin2016a} \\
\hline
\end{tabularx}
\end{table}

\begin{table}[H]
\centering
\small % or \scriptsize for even smaller
\caption{Parameters for Component A, B and C of Hip72217}
\label{tab:parameters}
\begin{tabularx}{\textwidth}{l c *{3}{>{\centering\arraybackslash}X}}
\hline
\textbf{Parameters} & \textbf{Units} & \textbf{Component A} & \textbf{Component B} & \textbf{Component C} \\
\hline

$M_{v} \pm \sigma M_{v}$     & (mag)       & 4.03 $\pm$  0.02   & 4.29 $\pm$  0.03   & 13.79 $\pm$ 0.2 \\
$M_{bol} \pm \sigma M_{bol}$ & (mag)       & 3.659 $\pm$ 0.01   & 3.888 $\pm$ 0.02   & 10.01 $\pm$ 0.1  \\
$L \pm \sigma L$             & ($L_{\odot}$)  & 2.73 $\pm$ 0.19   & 2.21 $\pm$ 0.14   & 0.007 $\pm$ 0.0002 \\
$T_{\mathrm{eff}} \pm \sigma T_{\mathrm{eff}}$ & (K) & 6125 $\pm$ 50 & 5950 $\pm$ 50 & 3250 $\pm$ 50 \\
$R \pm \sigma R$             & ($R_{\odot}$)  & 1.47 $\pm$ 0.04   & 1.39 $\pm$ 0.03   & 0.28 $\pm$ 0.01 \\
$M \pm \sigma M$             & ($M_{\odot}$)  & 1.14 $\pm$ 0.15   & 1.12 $\pm$ 0.14   & 0.17 $\pm$ 0.01 \\
$\log(g) \pm \sigma \log(g)$ & (cm\,s$^{-2}$) & 4.14 $\pm$ 0.11   & 4.20 $\pm$ 0.12   & 4.76 $\pm$ 0.17 \\
\hline
\end{tabularx}
\end{table}

\begin{figure}[H]
    \centering
    \includegraphics[width=0.9\textwidth]{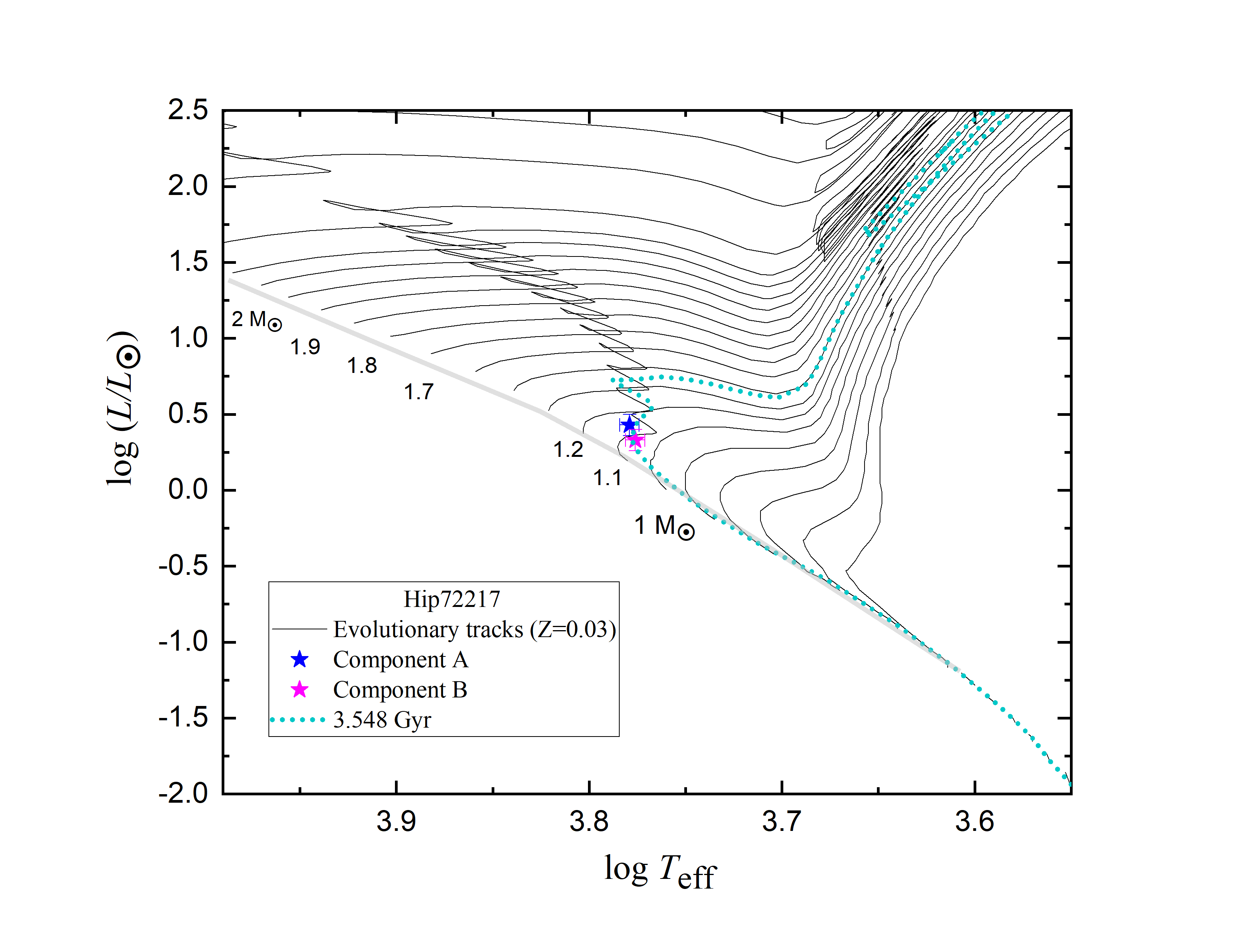}
    \caption{Component A and B of HIP~72217 placed on the Evolutionary Track and Isochrone of $Z=0.03$ taken from Girardi \cite{2000A&AS..141..371G, 2000yCat..41410371G}. The HR diagram shows a plot of $\log$ of effective temperature plotted against $\log(L/L_{\odot})$, the logarithm of the ratio of the stellar luminosity to the Solar Luminosity }
    \label{fig:evo_tracks}
\end{figure}

The positions of the components of HIP 72217 on the Hertzsprung-Russell diagram (HRD) are shown in Fig.~\ref{fig:evo_tracks} with the evolutionary tracks corresponding to the metallicity \(Z = 0.03\). The parts are as follows: \textbf{blue} stars refer to the primary star, while \textbf{magenta} stars refer to the secondary star.

The evolutionary tracks, which are derived from model stellar evolution, give an approximate age of the system by identifying the position of each star on the corresponding track. When compared with the nearest isochrone (\textbf{dotted cyan line}), the age of the system can be estimated to be \textbf{3.548 Gyr}.\\

As it can be seen from their positions on the diagram, \textbf{Component A} and \textbf{Component B} is somewhat post-MS and possesses some features of a subgiant. The estimated masses of the components based on the tracks are \(M_A \approx 1.14 M_{\odot}\) and \(M_B \approx 1.12 M_{\odot}\) which confirms the classification of the primary and secondary components as subgiants.\\

The results obtained from both the evolutionary tracks and SED fitting methods converge on a self-consistent model for the system, reinforcing the classification of HIP 72217s AB binary sub-system as subgiants. The mass and age estimations from the HRD align well with those inferred.\\

The derived age of $3.548 \pm 0.3~\mathrm{Gyr}$ for this system provides significant value when researchers compare it to other well-studied subgiant binaries. Theoretical predictions match the observed age range for stars in the solar-type mass range from $1.1\,M_{\odot}$ to $1.2\,M_{\odot}$ that move off the main sequence. Research shows that subgiant binaries with matching mass and metallicity characteristics normally have ages between $2.5$ and $6.5~\mathrm{Gyr}$~\cite{Torres_2009}. Asteroseismic and evolutionary modeling~\cite{refId0} determined that the $\alpha$~Centauri system, with $1.10\,M_{\odot}$ and $0.91\,M_{\odot}$ components, has an age of $6.5 \pm 1.0~\mathrm{Gyr}$. The evolutionary stage of HIP~72217 can be determined as early subgiant based on its proximity to the $3.5~\mathrm{Gyr}$ isochrone in the Hertzsprung–Russell diagram using $Z = 0.03$~\cite{Girardi2000a}. This suggests that both components A and B have only recently departed from the main sequence, marking the beginning of the hydrogen shell-burning phase. Our age determination through stellar evolutionary tracks and the subgiant system age distribution in published literature shows good agreement, which validates both the modeling approaches and the derived stellar parameters.

%\begin{figure}[H]
%\centering
%\includegraphics[width=0.7 \linewidth]{}
%\caption{}
%\label{Comparing Radii}
%\end{figure}

\subsection{Orbital Solution}
The HIP~72217 system's orbital solution was computed using Tokovinin's ORBITX code. The refined orbital elements and data from Mason(2010) are shown in Table~\ref{tab:measurements}. The table contains the final orbital elements derived from HIP 72217. The updated observational data combined with improved fitting methods delivers more precise dynamical mass estimates that are essential for stellar characterization. The Mason (2010) study provided initial orbital parameters for the binary star system HIP 72217, but the results in this work  indicates a more rigorous fit, due to  newer observational data points.

The total dynamical mass of the system was determined using Kepler's Third Law~\ref{eq:kepler_third_law} with a newly suggested parallax of $\pi = 25.65$\,mas. This adjustment, based on the system's observed properties, yielded a dynamical mass of $M_{\text{dyn}} = 2.26\,M_{\odot}$ which is in excellent agreement with the total mass obtained from Al-Wardat's method ($M_A + M_B = 2.26 \pm 0.29\,M_{\odot}$). The consistency between these independent methods validates both the orbital solution and the stellar parameter determination, while resolving discrepancies in previous mass estimates.

\begin{figure}[H]
    \centering
    \includegraphics[width=0.85\linewidth]{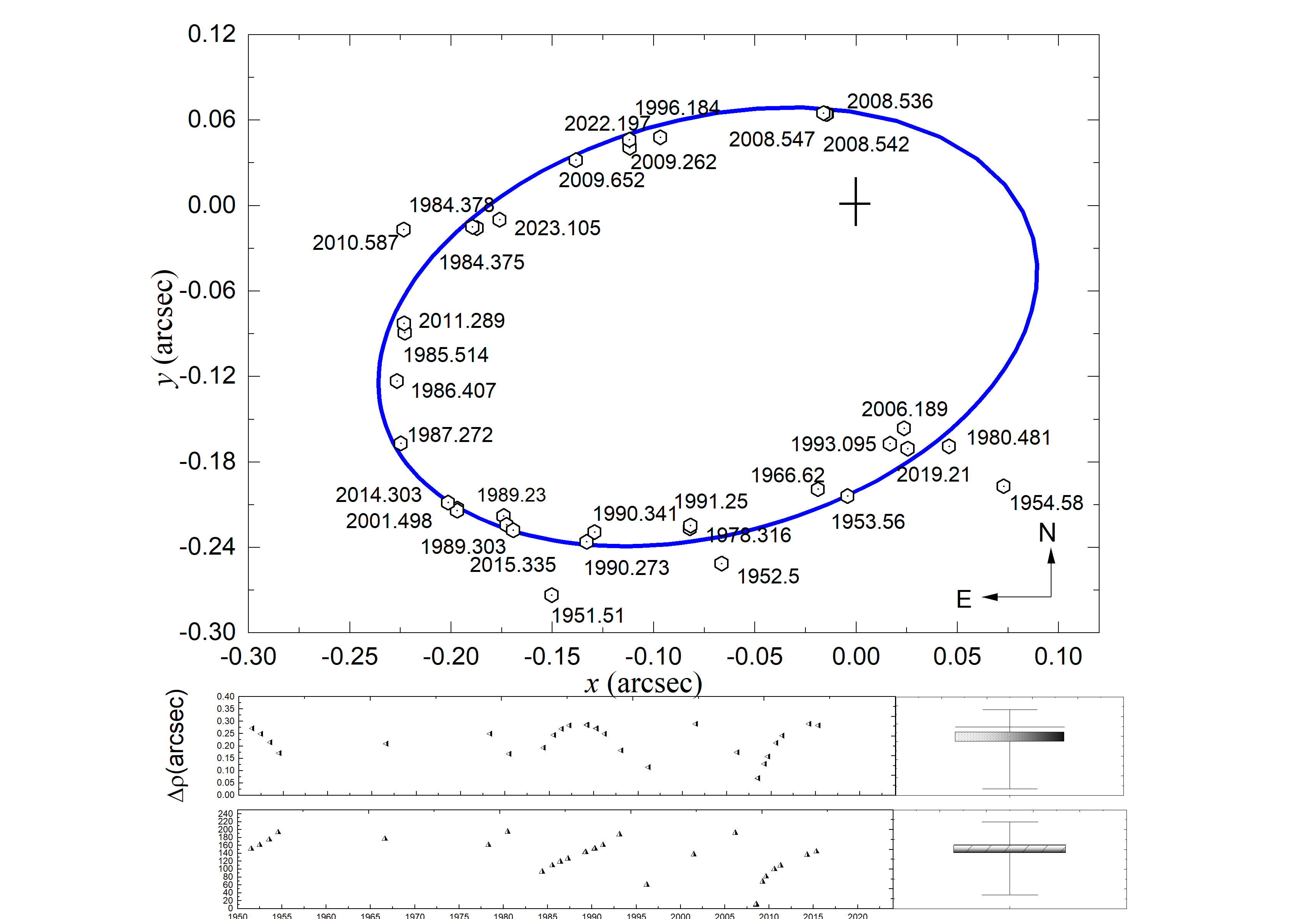}
    \caption{Orbit of HIP72217 (solid curve) with observations (circles; 1951--2023). The bottom left panel displays fit residuals that indicate the differences between observed and modeled angular separation ($\Delta\rho$) and position angle ($\Delta\theta$) measurements of the orbit. The bottom right panel shows a box plot that summarizes data distribution by displaying minimum and maximum values, first ($Q1$) and third ($Q3$) quartiles, median, and any detected outliers.}
    \label{fig:Orbit}
\end{figure}

The residuals of the orbital fit demonstrate high precision (RMS $\theta = 2.12^\circ$, RMS $\rho = 0.0127$\,arcsec). Fig.\ref{fig:Orbit} shows the computed orbit's excellent agreement with historical measurements from Table~\ref{tab:orbit}, further confirming the solution.

\begin{table}[H]
\centering
\caption{ Observational points and measurements with sources}
\label{tab:measurements}
\begin{tabular}{ccccc}
\hline
Date & $\theta$ (deg) & $\sigma_\theta$ & $\rho$ (deg) & Source \\ 
\hline
1951.510 & 151.2 & 0.312 & 0.001 & Fin1951b J \\
1952.500 & 165.2 & 0.260 & 0.001 & Fin1953d J \\
1953.560 & 178.8 & 0.204 & 0.011 & Fin1954c J \\
1954.580 & 200.3 & 0.210 & 0.001 & Fin1954c J \\
1966.620 & 174.6 & 0.200 & 0.001 & Fin1967a J \\
1978.316 & 160.1 & 0.241 & 0.001 & McA1984b Sc \\
1980.481 & 195.2 & 0.175 & 0.001 & McA1983 Sc \\
1984.375 & 94.8 & 0.188 & 0.001 & McA1987b Sc \\
1984.378 & 94.5 & 0.190 & 0.001 & McA1987b Sc \\
1985.514 & 111.8 & 0.240 & 0.001 & McA1987a Sc \\
1986.407 & 118.5 & 0.258 & 0.001 & McA1989 Sc \\
1987.272 & 126.6 & 0.280 & 0.001 & McA1989 Sc \\
1989.230 & 141.4 & 0.279 & 0.001 & McA1990 Sc \\
1989.303 & 142.4 & 0.283 & 0.001 & McA1990 Sc \\
1990.273 & 150.6 & 0.271 & 0.001 & Hrt1992b Sc \\
1990.341 & 150.6 & 0.263 & 0.001 & Hrt1993 Sc \\
1991.250 & 160.0 & 0.239 & 0.001 & HIP1997a Hh \\
1993.095 & 185.7 & 0.168 & 0.001 & Hrt2000a Sc \\
1996.184 & 63.6 & 0.108 & 0.001 & Hrt2000a Sc \\
2001.498 & 137.2 & 0.290 & 0.001 & Hor2008 S \\
2001.498 & 137.4 & 0.291 & 0.001 & Hor2008 S \\
2006.189 & 188.6 & 0.158 & 0.001 & Msn2010c Su \\
2008.536 & 12.6 & 0.066 & 0.001 & Tok2010 S \\
2008.542 & 12.9 & 0.066 & 0.001 & Tok2010 S \\
2008.547 & 13.9 & 0.067 & 0.001 & Tok2010 S \\
2009.262 & 70.0 & 0.119 & 0.001 & Tok2010 S \\
2009.652 & 77.0 & 0.142 & 0.003 & Rch2010 O \\
2010.587 & 94.3 & 0.224 & 0.001 & Msn2011d Su \\
2011.289 & 110.3 & 0.238 & 0.001 & Tok2012d St \\
2014.303 & 136.0 & 0.290 & 0.001 & Tok2015c St \\
2015.335 & 143.4 & 0.284 & 0.001 & Tok2016a St \\
2019.210 & 191.4 & 0.156 & 0.010 & Tok2020 \\
2022.197 & 70.8 & 0.120 & 0.010 & Msn2023 \\ 
2023.105 & 95.8 & 0.184 & 0.001 & Tok2024\\
\hline
\end{tabular}
\end{table}

\begin{table}[ht]
\centering
\caption{Refined orbital elements and residuals for HIP 72217 (with comparison to Mason(2010).}
\label{tab:orbit}
\begin{tabular}{lllll}
\hline
\textbf{Parameter}       & \textbf{Unit} & \textbf{This Work}       & \textbf{Mason(2010)\cite{Mason_2010}}  \\
\hline
Period \( P \)           & yr            & \( 12.917 \pm 0.0012 \)   & \( 12.929 \pm 0.021 \)      \\
Epoch \( T \)            & yr            & \( 1995.2607 \pm 0.0022 \)   & \( 1995.2490 \pm 0.0550 \)  \\
Eccentricity \( e \)     & --            & \( 0.6319 \pm 0.0012 \)      & \( 0.6428 \pm 0.0051 \)     \\
Semi-major axis \( a \)  & arcsec        & \( 0.1854 \pm 0.0003 \)      & \( 0.1814 \pm 0.0021 \)     \\
\(\Omega\)               & deg           & \( 277.42 \pm 0.96 \)        & \( 281.9 \pm 4.10 \)        \\
\(\omega\)               & deg           & \( 44.42 \pm 0.91 \)         & \( 39.5 \pm 4.70 \)         \\
Inclination \( i \)      & deg           & \( 25.29 \pm 0.31 \)         & \( 25.90 \pm 2.60 \)        \\
\(K_1\)                  & km/s          & \( 7.06 \pm 0.00 \)          & \( 7.06 \pm 0.00 \)         \\
\(K_2\)                  & km/s          & \( 7.32 \pm 0.00 \)          & \( 7.32 \pm 0.00 \)         \\
\(V_0\)                  & km/s          & \( -2.00 \pm 0.00 \)         & \( -2.00 \pm 0.00 \)        \\
\hline
RMS \( \theta \)         & deg           & \( 2.12 \)                   & --                          \\
RMS \( \rho \)           & arcsec        & \( 0.0127 \)                 & --                          \\
\(\chi^2\)               & --            & \( 6941.74 \)                & --                          \\
\hline 
\end{tabular}
\end{table}

\section{\textbf{Stability and Habitability of HD 72217}}\label{sec:hd72217}

\subsection{Stability Analysis}

Orbital stability is inherently complex due to numerous influencing factors, including initial conditions, mathematical frameworks, and physical constants \cite{szebehely1984review}. In the context of binary star systems, an orbit is typically considered stable if the primary orbital parameters, eccentricity, semimajor axis, and inclination, remain relatively unchanged over extensive periods.

Habitability, as defined by \cite{cockell2016habitability}, refers to the ability of an environment to support the metabolic processes of at least one known organism, thus allowing its survival, growth, and reproduction.

This study employs the empirical stability equations derived by \cite{holman1999long} to delineate the stable orbital zones around the HD 72217 binary system. Two orbital types are examined: circumstellar (S-type), where a planet orbits a single star, and circumbinary (P-type), where a planet orbits both stars. Critical stability boundaries for these scenarios depend primarily on the binary's semi-major axis, mass ratio, and eccentricity.

To assess orbital stability within the HD 72217 system, we apply the empirical expressions of Holman and Wiegert for S-type and P-type orbits. The critical semi-major axis for S-type orbits is given by:

\begin{align*}
a_{s} = a\, \big[ & (0.464 \pm 0.006) + (-0.380 \pm 0.010)\,\mu + (-0.631 \pm 0.034)\,e \\
                  & + (0.586 \pm 0.061)\,\mu e + (0.150 \pm 0.041)\,e^2 + (-0.198 \pm 0.047)\,\mu e^2 \big]
\end{align*}

For P-type orbits, the critical semi-major axis is:

\begin{align*}
a_{p} = a\, \big[ & (1.60 \pm 0.04) + (4.12 \pm 0.09)\,\mu + (5.10 \pm 0.05)\,e + (-4.27 \pm 0.17)\,\mu e \\
                  & + (-2.22 \pm 0.11)\,e^{2} + (-5.09 \pm 0.11)\,\mu^{2} + (4.61 \pm 0.36)\,\mu^{2}\,e^{2} \big]
\end{align*}

Here, \(a\) is the binary semi-major axis, \(e\) is the binary eccentricity, and \(\mu = \mathcal{M}_{1} / (\mathcal{M}_{1}+\mathcal{M}_{2})\) is the mass ratio of the primary to the total mass.

\subsection{Habitable Zone Calculation}

Habitable Zone (HZ) distances around stars are derived following the approach outlined by \cite{2013ApJ...770...82K}. The procedure begins by calculating the effective stellar flux, adjusted for stellar temperature differences compared to the Sun. The effective flux is calculated via:

\begin{equation}
S_\text{eff} = S_{\text{eff}\odot} + a\Delta T + b(\Delta T)^2 + c(\Delta T)^3 + d(\Delta T)^4
\label{eq:Seff_hd72217}
\end{equation}

where $\Delta T = T_\text{star} - T_\odot$, and $a$, $b$, $c$, and $d$ are empirically determined coefficients.

The corresponding habitable zone distance is:

\begin{equation}
d_\text{HZ} = \sqrt{\frac{L / L_\odot}{S_\text{eff}}},
\label{eq:HZradius_hd72217}
\end{equation}

where $L / L_\odot$ is the star's luminosity relative to the Sun.
Each star within the HD 72217 system has unique habitable zones determined individually. Table \ref{tab:hz-radii_hd72217} summarizes the stellar flux and resulting HZ distances for different climate scenarios.

\subsection{Habitable Zone Distances}
To define the habitable zone (HZ) for HD 72217, Kopparapu et al. (2013) proposed a classification of climate scenarios. According to different planetary climate models, these scenarios represent theoretical thresholds that constrain the inner and outer edges of the HZ. According to the Recent Venus scenario, Venus would have lost its surface water due to solar irradiation at its inner limit. A planet reaches the Runaway Greenhouse Limit when it experiences uncontrollable greenhouse warming, making it uninhabitable. Water vapor saturation in the atmosphere in the Moist Greenhouse scenario could lead to hydrogen escape into space, resulting in increased greenhouse gas concentrations. Water can still exist at a distance of the Maximum Greenhouse limit even after carbon dioxide-induced warming has reached its maximum. In addition, the Early Mars scenario posits that Mars may have had liquid water under past solar conditions, which represents the most distant boundary of the habitable zone (HZ). Based on these established scenarios, we can calculate the stellar flux (Seff) and HZ distance for both the primary and secondary stars.
Table \ref{tab:stability_hz_ranges_hd72217} provides stability limits and HZ ranges for both circumstellar and circumbinary orbits in the HD 72217 system.

\begin{figure}[H]
\centering
\includegraphics[width=\linewidth]{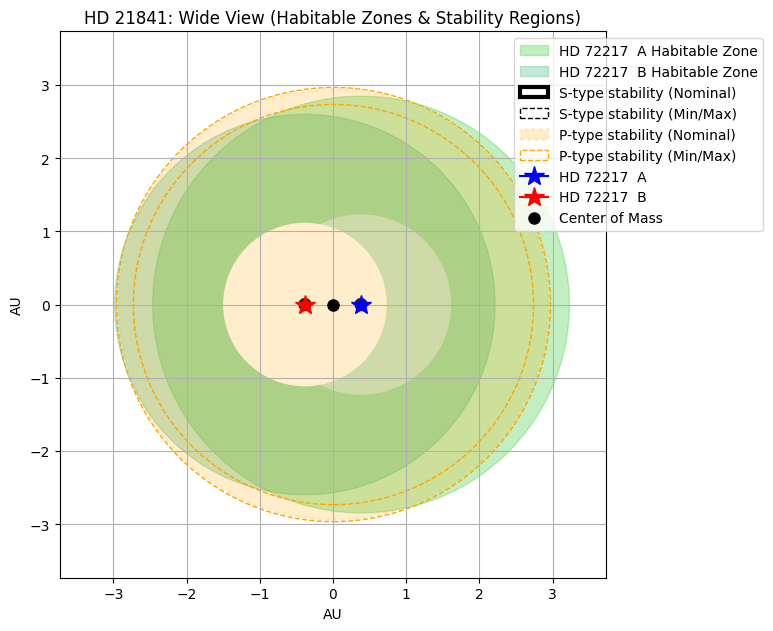}
\caption{Habitability (green area) and Stability (dashed line) for the HD72217 entire binary system. Using analytical models, this figure illustrates the potential habitable zones (green shaded areas) for HD 72217. It is entirely based on calculated criteria (Holman \& Wiegert, 1999; Kopparapu et al., 2013), without any direct observations. In this figure, no residuals are plotted since no empirical data are used. Within the system, it provides a conceptual map for evaluating potential planetary stability and habitability.}
\label{fig:HD72217_AB}
\end{figure}

\begin{figure}[H]
\centering
\includegraphics[width=\linewidth]{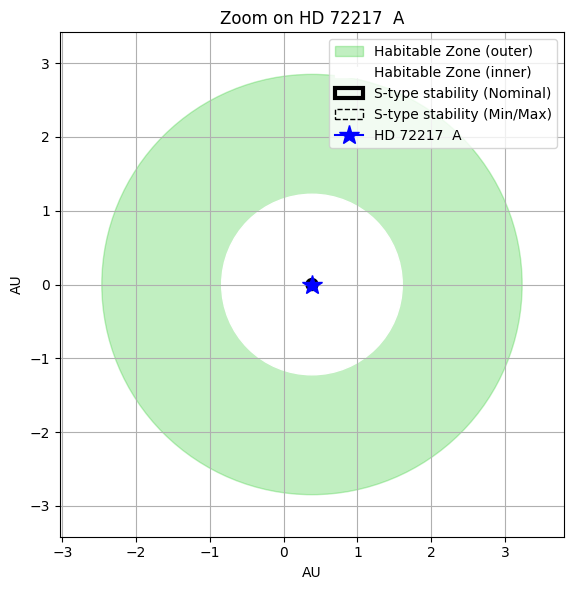}
\caption{Habitability and Stability zones around the primary star HD,72217 A.}
\label{fig:HD72217A}
\end{figure}

\begin{figure}[H]
\centering
\includegraphics[width=\linewidth]{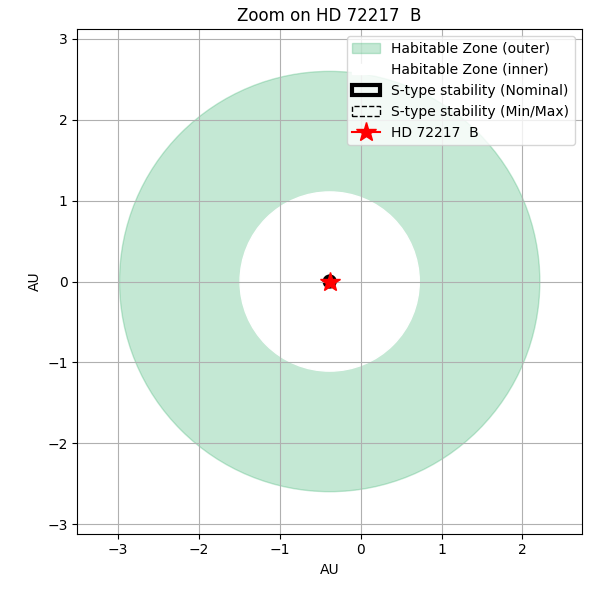}
\caption{Habitability and Stability zones around the secondary star HD,72217 B.}
\label{fig:HD72217B}
\end{figure}

Figure~\ref{fig:HD72217_AB} illustrates the overall habitability and stability regions for the complete HD 72217 binary system, while Figures \ref{fig:HD72217A} and \ref{fig:HD72217B} zoom in specifically on the regions around the primary and secondary stars, respectively. The shaded areas denote habitable zones, and dashed lines delineate stability boundaries where planetary orbits remain viable over long periods.
\begin{table}[ht]
\centering
\caption{This table summarizes the effective stellar fluxes (Seff) and the corresponding inner and outer boundaries of the habitable zone under five climate scenarios: Recent Venus, Runaway Greenhouse, Moist Greenhouse, Maximum Greenhouse, and Early Mars. Seff values and resulting HZ distances (in AU) are provided separately for both the primary and secondary components.}
\resizebox{\textwidth}{!}{%
\label{tab:hz-radii_hd72217}
\begin{tabular}{lcccccccccc}
\hline\hline
Scenario & $S_{\mathrm{eff,sun}}$ & $a$ & $b$ & $c$ & $d$ & $S_{\mathrm{eff,1}}$ & Primary HZ (AU) & $S_{\mathrm{eff,2}}$ & Secondary HZ (AU) \\
\hline
Recent Venus & 1.7753 & 0.000143 & 2.9875e-09 & -7.5702e-12 & -1.1635e-15 & 1.824718 & 1.223160 & 1.799685 & 1.108148 \\
Runaway Greenhouse & 1.0512 & 0.000132 & 1.5418e-08 & -7.9895e-12 & -1.8328e-15 & 1.098366 & 1.576550 & 1.074116 & 1.434401 \\
Moist Greenhouse & 1.0140 & 0.000082 & 1.7063e-09 & -4.3241e-12 & -6.6462e-16 & 1.042228 & 1.618452 & 1.027929 & 1.466272 \\
Maximum Greenhouse & 0.3438 & 0.000059 & 1.6558e-09 & -3.0045e-12 & -5.2983e-16 & 0.364201 & 2.737856 & 0.353853 & 2.499107 \\
Early Mars & 0.3179 & 0.000055 & 1.5313e-09 & -2.7786e-12 & -4.8997e-16 & 0.336768 & 2.847186 & 0.327197 & 2.598910 \\
\hline
\end{tabular}
}
\end{table}

\begin{table}[ht]
\centering
\caption{Stability Distances and Habitable Zone Ranges for HD\,72217 system}
\label{tab:stability_hz_ranges_hd72217}
\begin{tabular}{l c}
\hline\hline
\textbf{Parameter} & \textbf{Value (AU)} \\
\hline
S-type stability distance (nominal) & 0.062 \\
S-type stability distance (min)     & 0.003 \\
S-type stability distance (max)     & 0.067 \\
\hline
P-type stability distance (nominal) & 2.946 \\
P-type stability distance (min)     & 2.734 \\
P-type stability distance (max)     & 2.966 \\
\hline
HD 72217 A HZ range                 & 1.223 -- 2.847 \\
HD 72217 B HZ range                 & 1.108 -- 2.599 \\
\hline\hline
\end{tabular}
\end{table}

This detailed analysis is crucial for evaluating the potential habitability of planets within binary star systems, highlighting the strong influence of stellar properties on the extent and position of habitable zones.

Although our calculation defines theoretical limits to stability and habitability in the HD 72217 system, detection of planets in similar close binary systems is considerably challenging from observational viewpoints. Small stellar separations (about 0.182 mas in the case of HIP 72217 AB) pose challenges to radial velocity measurements because of combined gravitational and stellar activity-induced effects that can suppress the subtle effects of planets. Even if we use direct-imaging methods, the high brightness contrast and low angular stellar-planet separation pose major limitations that may be overcome if advanced methods like interferometric methods or high-contrast adaptive optics are employed with specialized and long-term observational missions in state-of-the-art telescopes.

\section{Conclusion}
In this work, we used Al-Wardat’s method for analyzing stellar systems and Tokovinin's dynamical method to study the triple system HIP 72217 and to determine the stellar parameters of its components. We found that two of the three components of the system are subgiant stars, which are A and B, while the third component, C is a red dwarf main sequence star.  The first component has a mass of \(M_A = 1.14 \pm 0.15 \, M_{\odot}\) and effective temperature \(T_{\text{eff},A} = 6125 \pm 50 \, \text{K}\), and the second component has mass \(M_B = 1.12 \pm 0.14 \, M_{\odot}\) and effective temperature \(T_{\text{eff},B} = 5950 \pm 50 \, \text{K}\).
While the C component has been assured to have a  mass of \(M_A = 0.17 \pm 0.04 \, M_{\odot}\)  with an effective temperature of \(T_{\text{eff},C} = 3250 \pm 50 \, \text{K}\). 

The derived AB system age of \(3.548 \, \text{Gyr}\) is consistent with the theoretical expectations for stars of mass $1.1$–$1.2$~$M_{\odot}$ on the main sequence and transitioning to the subgiant phase. Such evolutionary tracks agree with the evolutionary tracks from \cite{Girardi2000a} for Z = 0.03, which would have subgiant behaviour for masses around this age. Similar mass systems were reported to have subgiant binary ages of \(2.5 -4.5 \, \text{Gyr}\) in comparative studies such as \cite{Torres_2009} and \cite{Yousef2021} which supports the validity of our result.  Meanwhile, the age of the component C can't be assured since it lies at the bottom of the main sequence.
The orbital solution for HIP~72217, derived using Tokovinin's ORBITX code, demonstrates excellent consistency between dynamical and theoretical mass estimates. A revised parallax $\pi = 25.65~\mathrm{mas}$ yields a dynamical mass of $2.26\,M_{\odot}$, in precise agreement with Al\mbox{-}Wardat's method ($2.26 \pm 0.29\,M_{\odot}$), resolving prior discrepancies. The fit shows high precision, with low residuals (RMS $\theta = 2.12^{\circ}$, RMS $\rho = 0.0127''$), and the computed orbit closely matches historical observations. This robust agreement validates both the orbital parameters and the system's stellar properties.

This work includes a comprehensive analysis of the orbital stability of HD~72217 as well as the potential habitability of the system. We computed the critical semimajor axes for circumstellar (S-type) and circumbinary (P-type) orbital stability using the empirical criteria developed by Holman and Wiegert~\cite{holman1999long}. According to the results, S-type stable zones lie between $0.003$ and $0.067~\mathrm{AU}$, with a nominal boundary at $0.062~\mathrm{AU}$, while P-type stable zones lie beyond $2.734~\mathrm{AU}$, peaking at $2.946~\mathrm{AU}$. According to Kopparapu et~al.~\cite{2013ApJ...770...82K}, we used a stellar flux model to assess habitability, considering variations in effective temperature and luminosity. Therefore, the habitable zone of the primary star ranges from $1.223$ to $2.847~\mathrm{AU}$, while the habitable zone of the secondary extends from $1.108$ to $2.599~\mathrm{AU}$. These results show that the two components contribute distinct but overlapping habitable zones, collectively defining an area suitable for circumbinary life. Consequently, stellar characteristics are crucial to the development of a life-friendly environment. Such complex binary configurations require future observational campaigns and refined dynamical models to identify stable, life-sustaining exoplanets.

This analysis provides a valuable benchmark for testing and refining existing evolutionary models, particularly in the subgiant phase—a critical transitional stage between the main sequence and the red-giant branch. The consistent results with theoretical isochrones strengthen the reliability of current evolutionary tracks. These findings enhance our understanding of internal stellar processes such as hydrogen-shell burning and envelope expansion, and they contribute to improving the precision of stellar-population synthesis models. Ultimately, the detailed study of subgiant binaries like HD~72217 bridges the gap between observational astrophysics and theoretical stellar physics, advancing both fields.

\section{Acknowledgment}
%\begin{acknowledgements}
	%\section{Acknowledgments}
 
  This study employed a range of resources and tools, including SAO/NASA, the SIMBAD database, the Fourth Catalog of Interferometric Measurements of Binary Stars, IPAC data systems. Furthermore, it utilized the MCMC ORBIT code and codes of Al-Wardat's method for analyzing stellar systems, including its sub-codes for spectrophotometric calculations and its technique for estimating parallaxes, masses and metallicities.  

%\end{acknowledgements}
\bibliographystyle{unsrt}
\bibliography{bibtex}

\end{document}